\documentclass[prl,twocolumn,draft,amsmath,showpacs]{revtex4}
\usepackage{graphics}

\begin{document}

\input epsf

\title {Field-dependent thermal transport in the Haldane chain compound NENP}

\author{ A. V. Sologubenko,$^1$  T. Lorenz,$^1$ J. A. Mydosh,$^1$ 
A. Rosch,$^2$
K. C. Shortsleeves,$^3$  and M. M. Turnbull$^3$}

\affiliation{$^1$II. Physikalisches Institut, Universit\"{a}t zu K\"{o}ln, 50937 K\"{o}ln, Germany}
\affiliation{$^2$Institut f\"{u}r Theoretische Physik, Universit\"{a}t zu
K\"{o}ln, 50937 K\"{o}ln, Germany}
\affiliation{$^3$Carlson School of Chemistry and Biochemistry, Clark University,
Worcester, MA 01610, USA}

\begin{abstract}
  We present a study of the magnetic field-dependent thermal transport in the spin $S=1$ chain material Ni(C$_2$H$_8$N$_2$)$_2$NO$_2$(ClO$_4$) (NENP).  The measured thermal conductivity is found to be very sensitive to the field-induced changes in the spin excitation spectrum. The magnetic contribution to the total heat conductivity is analyzed in terms of a quasiparticle model, and we obtain a temperature and momentum independent mean free path.  This implies that the motion of quasiparticles is effectively three dimensional despite the tiny interchain coupling.
\end{abstract}
\maketitle

The recent theoretical interest concerning heat transport in one
dimensional (1D) spin systems 
(see review article \cite{Zotos04_BookCh}  and the most recent papers \cite{Jung07,Boulat07,Savin07,Louis06,HeidrichMeisner05,Rozhkov05,Sakai05})
was
greatly stimulated by observations of a large spin thermal
conductivity $\kappa_s$ in  two-leg Heisenberg $S=1/2$ ladder
compounds (La,Sr,Ca)$_{14}$Cu$_{24}$O$_{41}$ at high temperatures,
with the mean free path of spin excitations $l$ reaching 3000 \AA~
\cite{Sologubenko00_lad,Hess01,Kudo01_Spi}.  However, experiments on
several Heisenberg $S=1$ chain compounds
\cite{Sologubenko03_AVPS,Kordonis06,Kawamata07} provided evidence for
considerably lower $\kappa_s$, with $l \le 60$~\AA . This is
surprising because the $S=1$ chain model and the $S=1/2$ ladder model
are essentially equivalent \cite{White96}.  Both adopt a spin liquid
state with exponentially decaying correlations and an energy gap in
the spin excitation spectrum.

The spin thermal conductivity $\kappa_s$ at a finite frequency
$\omega$ can be decomposed into a singular and a regular part
\begin{equation}
\label{ReKappa}
{\rm Re}\,\kappa_s(\omega) = D_{\rm th} \delta(\omega) + \kappa_{\rm reg}(\omega),
\end{equation}
where $D_{\rm th}$ is the thermal Drude weight.  In our experiment, we
measure the $dc$ conductivity ${\rm Re }\,\kappa_s(\omega=0)$. 
For integrable models and continuum field theories, heat
transport is ballistic even at finite temperature $T$, $D_{\rm th}>0$,
as conservation laws prohibit the decay of the heat current. 
Both extrinsic sources of scattering,  such as defects and phonons,
and intrinsic spin-spin interactions which are not integrable, render the heat
conductivity finite \cite{Zotos04_BookCh,Jung07,Boulat07,Savin07,Louis06,HeidrichMeisner05,Rozhkov05,Orignac03,Karadamoglou04}.

Measurements of the thermal conductivity in external magnetic fields,
which can strongly modify the spin excitation spectrum, offer detailed
information on scattering mechanisms limiting $\kappa_s$.  However,
magnetic fields typically available are too weak to noticeably
influence the spectrum of spin excitations in the previously
investigated \cite{Sologubenko03_AVPS,Kordonis06,Kawamata07} $S=1$ chain
materials AgVP$_2$S$_6$ and Y$_2$BaNiO$_5$ with strong intrachain
exchange $J>250 {\rm ~K}$.

In this Letter, we present results on field-dependent heat transport in
one of the model low-$J$ $S=1$ chain materials,
Ni(C$_2$H$_8$N$_2$)$_2$NO$_2$(ClO$_4$), {\it viz}. NENP.  The mean free
path of the spin excitations, evaluated from our data, is large and
temperature-independent allowing us to identify the most relevant
scattering mechanism. 
The heat transport at low temperatures is determined by rare defects, cutting the spin chains into segments,
and not by the intrinsic interactions.

NENP crystallizes in the orthorhombic $Pnma$ space group with lattice
parameters $a$=15.223~\AA, $b$=10.300~\AA, and $c$=8.295~\AA
~\cite{Meyer82}.  The $S=1$ spins of Ni$^{2+}$ form chains along the
$b$ axis with exchange constant $J \approx 43 {\rm ~K}$
\cite{Sieling00}, while the interaction $J'$ between the chains is
much weaker 
($J'/J\sim 8 \times 10^{-4}$ according to  
\cite{Regnault94}). 
Therefore, 
low-temperature 3D ordering is neither expected nor observed.
Neglecting the interchain interaction for the moment, the appropriate Hamiltonian is
\begin{eqnarray}\label{eHamiltonian}
H & = &  \sum_{i} \{ J{\bf S}^{i} {\bf S}^{i+1} + D (S_{z}^{i})^{2}  + E [ (S_{x}^{i})^{2} - (S_{y}^{i})^{2} ] \nonumber\\
 & + & \mu_B {\bf S}^i \cdot  {\bf g B} \} ,
\end{eqnarray}
where $D$ and $E$ are single-ion anisotropy constants, $B$ is the
magnetic field, and ${\bf g}$ is the gyromagnetic tensor.  In an ideal
isotropic antiferromagnetic (AFM) $S=1$ chain, the excitations are
triply degenerate with a gap $\Delta=0.41 J$ at the AFM wavevector
$k_{\rm AF} = \pi/d$, where $d=b/2$ is the distance between
neighboring spins along the chain. In NENP, the strong planar
anisotropy and weak orthorhombic anisotropy ($D/J=0.2$ and
$E/J\approx0.01$ \cite{Meyer82,Delica91}) split $\Delta$ into three
gaps $\Delta_{i}$ ($i = 1,2,3$) with the zero-field values $\Delta^0_1
\approx 29 {\rm ~K}$, $\Delta^0_2=14.3 {\rm ~K}$, and $\Delta^0_3=12.2
{\rm ~K}$ \cite{Regnault94}.  With increasing $B \parallel b$,
$\Delta_1$ stays constant, $\Delta_2$ increases, and $\Delta_3$
decreases such that it should close at the critical field $B_c \approx
10$~T and the system should enter a gapless Luttinger liquid (LL)
state for $B>B_c$ \cite{Konik02}.  This does not happen, however,
because the chemical environment of every second Ni atoms along the
chain is oriented in a different direction \cite{Chiba91}. This
alternating tilting introduces an additional term $\sum_{i}(-1)^i
\mu_B {\bf S}^i \cdot {\bf g}_{\rm st} {\bf B}$ in the Hamiltonian
Eq.~(\ref{eHamiltonian}), where the staggered transverse field $B_{\rm
  st}=| {\bf g}_{\rm st} {\bf B}|$ is proportional to the homogeneous
field $B$. As a consequence, the gap remains finite at $B=B_c$ and
increases above $B_c$.

The crystals of NENP used for our experiments were grown as described
in Ref.~\cite{Long07}.  One crystal of dimensions
1.1$\times$3$\times$0.5 ~mm$^3$ with the longest dimension along the
$b$ axis was used for the measurements of the thermal conductivity
along the spin chains. A sample of dimensions
1.8$\times$0.8$\times$1.4 ~mm$^3$ for measurements perpendicular to
the chains was cut from another crystal. For the thermal conductivity
measurements, we employed a standard steady state method with the same
arrangement of the thermometers and heater as described in
Ref.~\cite{Sologubenko07}.  The experiments were performed in the
temperature range between 0.3 and 50~K in magnetic fields up to 16~T
applied parallel to the chain direction.
 
\begin{figure}[tb]
   \begin{center}
    \leavevmode
    \epsfxsize=0.9\columnwidth \epsfbox {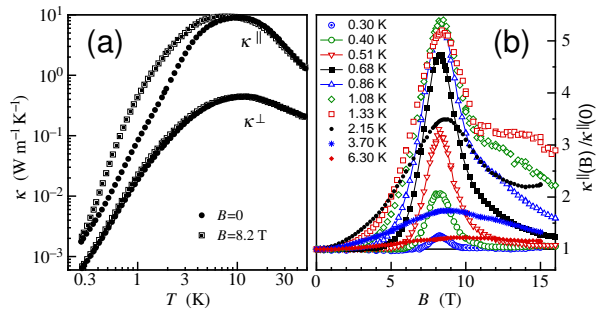}
     \caption{
     (a) Thermal conductivity of NENP parallel and perpendicular to the spin chains as 
     a function of temperature at  $B=0$ and $B=8.2 {\rm ~T}$. 
     (b) The relative change of the thermal conductivity of NENP parallel to the spin chains as 
     a function of magnetic field at several constant temperatures. 
    }
\label{KK0vsH}
\end{center}
\end{figure}

The thermal conductivity $\kappa(T)$ both parallel
($\kappa^{\parallel}$) and perpendicular ($\kappa^{\perp}$) to the
chain direction in zero field and in $B=8.2 {\rm ~T}$ are shown in
Fig.~\ref{KK0vsH}(a).  The relative changes of $\kappa^{\parallel}$ as
a function of magnetic field $\kappa^{\parallel}(B) /
\kappa^{\parallel}(0)$ at several constant temperatures are displayed
in Fig.~\ref{KK0vsH}(b).  The striking observation is that a magnetic
field leads to a strong enhancement of the thermal conductivity up to
5 times its zero-field value.  This strong enhancement is restricted
to the direction parallel to the chains.  The small increase of
$\kappa^{\perp}(B)$ can easily be attributed to a less than 1$^{\circ}$ deviation of the heat flow
direction from being exactly perpendicular to the $b$ axis.

The total thermal conductivity of a magnetic insulator can be
represented as $\kappa = \kappa_{\rm ph} + \kappa_s$, where the two
terms on the right-hand side correspond, respectively, to the phononic
and magnetic contributions to the heat transport (possible spin-phonon drag contributions are
included in $\kappa_s$). 
The spin excitation
spectrum at $B=0$ is gapped with the smallest gap $\Delta_3=12.2 {\rm
  ~K}$, therefore at $T \ll \Delta_3$, both spin thermal conductivity
and phonon-spin scattering are negligible.  With increasing $B$ at a
constant temperature, $\Delta_{3} (B)$ decreases and the number of
thermally activated spin excitations increases.  Therefore,
$\kappa_s(B)$ should increase, while $\kappa_{\rm ph}(B)$ should
decrease because of the growing phonon scattering by spin excitations.
Because of the quasi-1D nature of the spin system in NENP, 
$\kappa_s^{\perp}$ is small and $\kappa^{\perp} \approx
\kappa^{\perp}_{\rm ph}$. Thus, the observed negligible influence of
the magnetic field on $\kappa^{\perp}$ suggests that spin-phonon
scattering is weak. The increase of $\kappa^{\parallel}(B)$ in fields
$0 < B < B_c$ clearly demonstrates that all field-induced changes in
$\kappa^{\parallel}$ originate from $\kappa_s$.

A salient feature of the curves shown in Fig.~\ref{KK0vsH}(b) is that for all temperatures below about 1.5~K there is a region of low fields where $\kappa^{\parallel}(B)/\kappa^{\parallel}(0)$ are practically field-independent. This means that below about 1.5~K $\kappa^{\parallel}(B=0)$ is purely phononic. 
Therefore, by subtracting the zero-field values from $\kappa^{\parallel}$  at these temperatures  we obtain a good estimate of  $\kappa_s(B)$. 

\begin{figure}[tb]
   \begin{center}
    \leavevmode
    \epsfxsize=0.9\columnwidth \epsfbox {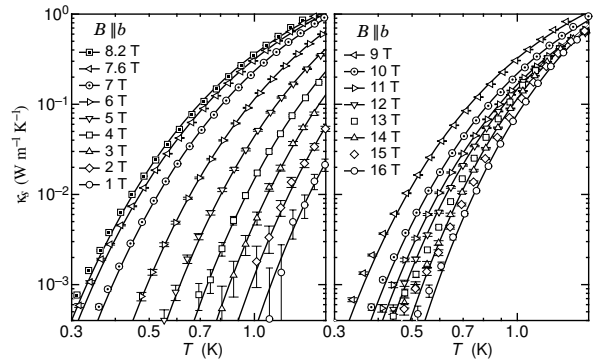}
    \caption{ Magnetic contribution to the thermal conductivity of
      NENP parallel to the spin chains as a function of temperature at
      several constant fields. The solid lines represent the
      calculated $\kappa_s$ (see text). The error bars arise from the
      uncertainty in subtraction of $\kappa_{\rm ph}$ from the total
      measured thermal conductivity.  }
\label{KmT}
\end{center}
\end{figure}

We have measured $\kappa^{\parallel}(T)$ below 1.5~K at several
constant fields.  The spin contribution $\kappa_s(B,T) \equiv
\kappa^{\parallel}(B,T) - \kappa^{\parallel}(0,T)$ is shown in
Fig.~\ref{KmT}.  For a quantitative analysis of the data, we consider
the heat transport associated with excitations from the singlet ground
state to the lowest triplet branch.  As the energy gaps for the other
two branches of the triplet either increase with $B$ or stay constant,
their contribution to the heat transport below $T \approx 1.5$~K can
be disregarded.  The interaction between chains is very weak in
comparison with the intrachain interaction; nevertheless, it still
leads to a dispersion perpendicular to the $b$ axis with a bandwidth
of about 2~K and 0.8~K along the $a$ and $c$ axes, respectively
\cite{Regnault94,Zaliznyak98}.
The dispersion  we use in order to analyze $\kappa_s$ in low
fields (hence the subscript "$lf$") 1~T $\leq B \leq 6$~T  is  given by
\begin{eqnarray}\label{FullDispersion}
\varepsilon_{lf}({\bf k}) & = & \{ [(\Delta_2^0 + \Delta_3^0)/2]^2 + V^2 (k_b d-\pi)^2 \nonumber\\
& + &  (\Delta E_a)^2[(1+\cos k_a a)/2]^2 \nonumber\\
& + &  (\Delta E_c)^2[(1+\cos k_c c)/2]^2 \}^{1/2} - g_b \mu_B B,
\end{eqnarray}
 with constants $\Delta_2^0 = 14.3$~K, $\Delta_3^0 = 12.2$~K, $\Delta
E_a=7.5$~K, and $\Delta E_c=5.0$~K taken from neutron scattering
experiments \cite{Regnault94,Zaliznyak98} and $g_b=2.15$ from an ESR
study \cite{Sieling00}.  The value of $V=2.49 J$, used in the
analysis, is predicted by theory \cite{Sorensen93} and is confirmed by
the inelastic neutron scattering measurements \cite{Ma92,Regnault94}.
Equation (\ref{FullDispersion}) describes correctly the linear-in-$B$
decrease of the energy gap at ${\bf q}=(0,\pi/d,0)$ between 1 and 6~T
observed in the ESR experiment \cite{Sieling00}, but does not account
for deviations at higher fields caused by the influence of the
aforementioned transverse staggered field.

Within the Boltzmann equation approximation, the spin thermal
conductivity parallel to the chain direction is represented by
\begin{equation}\label{eMFTKappa}
 \kappa_s =  \frac{n}{\hbar \pi^3}   l 
 \int\limits_{0}^{\pi/a} dk_a
 \int\limits_{0}^{\pi/c} dk_c
  \int\limits_{0}^{\pi/b}{\frac{df}{dT} \varepsilon({\bf k})    \frac{d\varepsilon({\bf k})}{dk_b}   dk_b}, 
\end{equation}
where $n=4$ is the number of spins in the unit cell of NENP, $f$ is
the distribution function, $\varepsilon({\bf k})$ is the energy, and
$l$ is the mean free path of spin excitations. In
Eq.~(\ref{eMFTKappa}), we use the Fermi distribution $f =
(\exp(\varepsilon/k_B T)+1)^{-1}$ (taking into account the
hard-core repulsion of the 1D excitations) \cite{Huang04}, but as the
temperature is
well below the gap, one can equally well use a Boltzmann- or Bose
distribution.

\begin{figure}[t]
   \begin{center}
    \leavevmode
    \epsfxsize=0.6\columnwidth \epsfbox {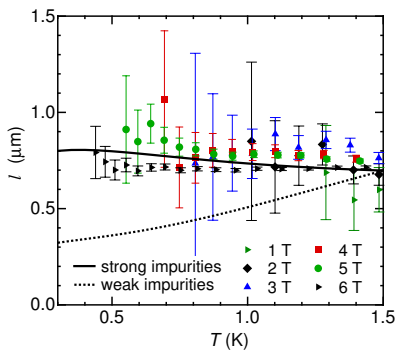}
    \caption{ Mean free path of spin excitations, calculated for
      1~T~$\leq B \leq$~6~T. The error bars correspond to those of
      $\kappa_s$ shown in Fig.~\ref{KmT}. An approximately constant
      mean free path can be obtained from a model (solid line) where strong impurities  effectively
      cut the spin chains into segments of length $l$. In contrast,
      the data cannot be described  (dashed line) by weak fluctuations
      of $J$.}
\label{Ell}
\end{center}
\end{figure}

In the field region 1~T~$\leq B \leq$~6~T, where $\varepsilon({\bf k})
= \varepsilon_{lf}({\bf k})$, the only unknown parameter in
Eq.~(\ref{eMFTKappa}) is $l$.  Thus, the mean free path, $l(B,T)$, as a function
of $B$ and $T$ can be extracted from the experimental data of
Fig.~\ref{KmT} and is shown in
Fig.~\ref{Ell}.  Within the experimental accuracy, the mean free path
is independent of both $T$ and $B$, with the
average value ${\bar l} = 0.75 \pm 0.1$~$\mu$m.  Remarkably, ${\bar
  l}$ for NENP is as large as the highest values of the mean free path
found for $S=1/2$ Heisenberg chains and
ladders \cite{Sologubenko07,Sologubenko03_Uni,Hess01}, where impurities
are the main source of scattering for spin excitations at low
$T$. Both for spin-spin and spin-phonon scattering, the mean
free path is expected to increase rapidly with decreasing $T$ \cite{Boulat07}. We therefore
conclude that, at least at $T<0.04 J$, scattering by defects and not the
intrinsic interactions determine $\kappa_s$. 
A mean free path due to the intrinsic processes is theoretically expected to be huge \cite{Jung07}
as Umklapp processes which relax momentum are exponentially supressed for $T \ll J$.

A field and temperature independent mean free path cannot be obtained
from a purely 1D model. For both weak and strong
impurities, one expects \cite{Sologubenko07} in one dimension a
momentum dependent mean free path proportional to the square of the
velocity, $l_k \propto v_k^2$, implying effectively a variation of $l$
linear with $T$. The situation is, however, completely different when
one takes the tiny three dimensional coupling between the chains into
account. If spin excitations can pass "strong" impurities, which
effectively cut the chains into segments, by hopping to the next chain
instead of tunneling through the defect, one naturally
obtains a mean free path given by the distance of the defects. We have
calculated the weak $T$ dependence of $l$ (solid line in
Fig.~\ref{Ell}) assuming a small density of local, infinitely strong
potential scatterers in a model defined by the dispersion relation
(\ref{FullDispersion}) for a gap $\Delta$ of $5$\,K (for $T\ll \Delta$ the
result is almost independent of $\Delta$). Such a
simple calculation to linear order in the density of defects is valid
as the interchain coupling $J'$ (or more precisly the bandwidth in
perpendicular direction) is much larger than the inverse of the time
needed by a spin excitation with energy $T$ to propagate to the next
defect, $J' > \sqrt{T J} d/l$.

The calculations for the mean free path dominated by 
strong impurities describe the data very well (solid line in Fig.~\ref{Ell}). 
This is not the case when one assumes that small fluctuations
$\delta J \lesssim T$ of $J$ dominate transport (dashed
line in Fig.~\ref{Ell}). 
In the latter case, the scattering rate is proportional to the density of state [obtained
again from Eq.~(\ref{FullDispersion})].

\begin{figure}[t]
   \begin{center}
    \leavevmode
    \epsfxsize=0.9\columnwidth \epsfbox {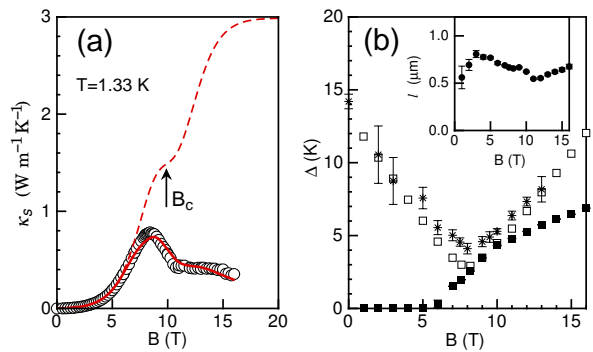}
    \caption{ (a) $\kappa_s(B)$ calculated for $T=1.33$~K with (solid
      line) and without (dashed line) consideration of the staggered
      transverse field. The open circles represent the experimental
      data.  (b)  The open squares are the minimum energy gap
      $\Delta_{\rm min}$ estimated in our analysis. For comparison,
      the energy gap obtained from specific heat measurements from
      Ref.~\cite{Kobayashi92} are show (stars). The solid squares are the fitted values for
      $\Delta_{\perp}$ (the vanishing of $\Delta_{\perp}$ below $6$\,T may be an artifact of the fitting procedure as $\Delta_{\rm
        min}$ depends very litte on $\Delta_{\perp}$ in this regime). Inset: The fit
      values of the mean free path of spin excitations. }
\label{Fit}
\end{center}
\end{figure}

The model we used for the low-field calculations fails to describe
$\kappa_s(B,T)$ at $B > 6$~T.  This is illustrated in
Fig.~\ref{Fit}(a), where the dashed line is $\kappa_s(B)$ calculated
for a constant temperature $T \ll J$ using
Eqs.~(\ref{FullDispersion}) and (\ref{eMFTKappa}) with $l = {\bar l}$.
The calculated $\kappa_s(B)$ shows an increase when $B$ approaches
the LL state from the gapped state, with a plateaulike feature at
$B_c$ broadened by the interchain interaction.  This is the expected
generic behavior for a 1D spin system near a quantum phase
transition from a gapped to a gapless state and has, indeed, been
observed for the $S=1/2$ chain compound CuPzN in \cite{Sologubenko07}.
The different behavior of NENP for high fields arises
obviously from the energy gap induced by the staggered field.

We have fitted Eq.~(\ref{eMFTKappa}) to the $\kappa_s(T)$ data
presented in Fig.~\ref{KmT} using a modified dispersion relation
$\varepsilon({\bf k}) = \sqrt{ \Delta_{\perp}^2 +
  \varepsilon_{lf}^2({\bf k}) }$, where $\varepsilon_{lf}({\bf k})$
is given by Eq.~(\ref{FullDispersion}) and the energy gap
$\Delta_{\perp}$ is induced by the staggered field. A similar
heuristic fitting formula has e.g. been used in
Ref.~\cite{Kenzelmann05} to describe the $S=1/2$ chain compound
CuCl$_2\cdot$2((CD$_3$)$_2$SO) where, similar to NENP, a staggered
field induced by a uniform magnetic field leads to a finite gap in the
spin excitation spectrum.  For each value of $B$
between 1~T and 16~T, Eq.~(\ref{eMFTKappa}) was fitted to the
experimental $\kappa_s(T)$ data, shown in Fig.~\ref{KmT}, with two free parameters $l$ and $\Delta_{\perp}$,  assuming that $l$ is $T$-independent for each $B$.
The resulting $\kappa_s(T)$ curves 
are shown in Fig.~\ref{KmT}.  The fit values for $\Delta_{\perp}(B)$ are shown by
solid squares in Fig.~\ref{Fit}(b), the data for $l(B)$ are presented
in the inset of Fig.~\ref{Fit}(b).  The solid line in
Fig.~\ref{Fit}(a) is calculated using these data for $l(B)$ and
$\Delta_{\perp}(B)$.  As shown in Fig.~\ref{Fit}(b), there is an
agreement between our data for the minimum energy gap $\Delta_{\rm
  min} \equiv [((\Delta_2^0 + \Delta_3^0)/2 - g_b \mu_B B )^2 +
\Delta_{\perp}^2]^{1/2}$ at ${\bf k}=(\pi/a,\pi/d,\pi/c)$ and the data
for the energy gap obtained from the specific heat measurements
\cite{Kobayashi92}.
 The essential result is that the mean free path $l$ remains close to its
low-field value ${\bar l} = 0.75$~ $\mu$m in the entire 1-16~T
field region.  This is again consistent with the notion of a
 mean free path determined by  strong impurities.

In summary, from our measurements of the anisotropic thermal
conductivity of the $S=1$ Haldane chain compound NENP, we identify a
large magnetic contribution along the spin chain direction. The mean
free path of spin excitations is orders of magnitude larger than
previously observed for other $S=1$ chain materials and of the same
order of magnitude as in the best $S=1/2$ chain and ladder compounds.
We have argued that the absence of a temperature and field dependence
of the mean free path can be explained by rare defects, which
effectively cut the spin chains into segments, in combination with a
tiny interchain coupling.  The measured values of spin thermal
conductivity may also serve as a lower limit for future theoretical
estimates of the intrinsic diffusive contribution to the heat
transport in $S=1$ AFM chains at low temperatures.

 \acknowledgments

We acknowledge useful discussions with A.~Altland, D.~I. Khomskii,
M. Garst, E.~Shimshoni, M.~Vojta and S.~Zvyagin. This work was
supported  by the DFG through SFB 608.

\end{document}